\documentclass{ws-procs975x65}
 \begin{document} 
\title{Merging Gravitation with Thermodynamics to Understand Cosmology}
\author{Andrew P. Lundgren}
\address{Albert Einstein Institut, Callinstr. 38,  
30167 Hannover, Germany\\
E-mail: andrew.lundgren@aei.mpg.de}
\author{Ruxandra Bondarescu$^*$} 
\address{ Institute for Theoretical Physics, Winterthurerstrasse 190, 
CH-8057, Zurich, Switzerland \\
E-mail: ruxandra@physik.uzh.ch}
\author{Mihai Bondarescu}
\address{108 Lewis Hall, University of Mississippi, Oxford, MS 38677, USA}
\address{Facultatea de Fizica, Universitatea de Vest, 
Blvd.~V.~Parvan 4, Timisoara 300223, Romania \\
E-mail: mihai7@gmail.com}
\begin{abstract}
 We discuss the evolution of the universe in the context of the second law of thermodynamics from its early stages to the far future. Cosmological observations suggest that most matter and radiation will be absorbed by the cosmological horizon. On the local scale, the matter that is not ejected from our supercluster will collapse to a supermassive black hole and then slowly evaporate.  The history of the universe is that of an approach to the equilibrium state of the gravitational field. \end{abstract}
\begin{center}
{\bf Essay Written for the Gravity Research Foundation 2013 Awards \\
for Essays on Gravitation. \\
Submitted on March 31st.}
\end{center}
\newpage
\section{Introduction}
Our universe started uniform and hot \cite{ref1} and became colder and clumpier as more and
more structure formed \cite{ref4}. Early on, radiation and later matter dominated the universe \cite{ref5}.
As the universe expanded, the concentrations of matter and radiation decreased, and dark energy
started to dominate the universe causing the accelerated expansion we observe today \cite{ref6,ref7}.
At first this seems to violate the second law of thermodynamics, that entropy never decreases, because the hot and uniform early universe appears to have high entropy and the clumpy, cold universe of today has low entropy. However, since gravitational entropy increases with clumpiness, the total entropy increases and the second law of thermodynamics is not violated. Three of the largest repositories of entropy in the current universe are gravitational: the cosmological event horizon, the supermassive black holes, and the cosmic gravitational wave background \cite{ref12,ref13}.

\section{The Far Future of the Local Universe}
In time, the universe will become larger and emptier. The cosmological horizon absorbs entropy from the matter and radiation that stream through it, and grows in response. Eventually, most matter and radiation will stream beyond the cosmological horizon leaving dark energy to completely dominate the universe on scales beyond $100$ Mpc. However, the local supercluster is decoupled from the expansion of the universe and will remain gravitationally bound \cite{ref8}.  Imagine now that the local supercluster collapses to a single supermassive black hole of $M \sim 10^{15} M_\odot$ \cite{ref16}, which has a Schwarzschild radius of $300$ light years and a Hawking temperature of $T_{\rm BH} \approx 1/(8 \pi M) \sim  10^{-23}$ K \cite{ref17, ref10}. Once the universe cools below this temperature, the black hole will start to evaporate. The cosmic microwave background cools beyond the black hole temperature in $10^{12}$ years and below the temperature of the cosmological horizon ($T_{\rm C} \approx 1/(2 \pi \alpha) \sim 10^{-30}$ K) in $1.4 \times 10^{12}$ years. Here $\alpha$ is the areal radius of the cosmological horizon if the universe were empty de Sitter. The timescales are close due to the exponential nature of the expansion of the universe during dark energy domination. 

The heat from the relatively hot black hole will then flow across the much colder cosmological horizon. This heat flow produces much more entropy than was lost by the black black hole \cite{ref17, ref10} $S_{BH} \approx 4 \pi M^2$. The total entropy of empty de Sitter with a non-rotating black hole added at its center is\cite{ref21,refour} $S_{\rm tot} \approx S_C = \pi (\alpha^2 - 2 \alpha M  + ....)$, where $\alpha$ is given by the cosmological constant $\Lambda = 3/\alpha^2$ or alternatively by the Hubble constant $H = 1/\alpha$. The total entropy $S_{\rm tot}$ increases as the mass of the black hole decreases due to the $- 2 \alpha M$ term, which is much larger than $S_{\rm BH}$ because $\alpha \gg M$.

Everything in the local universe is expected to either coalesce into supermassive black holes or be ejected \cite{ref9}. The supermassive black holes will then slowly evaporate due to Hawking radiation \cite{ref10}. The evaporation process clearly satisfies the second law of thermodynamics. Heat flows from hot (black hole) to cold (cosmological horizon) and $S_{\rm tot}$ increases. Since black hole evaporation is the dominant non-equilibrium process, black hole thermodynamics is the physics of the far future.

\section{The Entropy of Flat Minkowski Space}
Operationally, Minkowski space is indistinguishable from de Sitter with the cosmological constant going to zero. In the $\Lambda \to 0$ limit of de Sitter, temperature goes to zero and entropy diverges with the area of the cosmological horizon. In the literature, the entropy of flat space is sometimes taken to be zero \cite{ref19} because flat space has no horizon, although \cite{ref20} argues that the entropy is infinite. This confusion could arise due to the Weyl Curvature Hypothesis \cite{ref25}, which states that the difference in entropy between the initial and final states of the universe is related to the growth of the overall Weyl curvature. However, the de Sitter metric has zero Weyl curvature but extremely large entropy, so there cannot be a direct relation between Weyl curvature and gravitational entropy. 

We propose a more subtle relation between the Weyl curvature and the gravitational entropy. The Einstein equations determine the Riemann curvature locally by the stress-energy tensor, but the Weyl tensor is nearly free except for boundary conditions given by the Bianchi identities. The modes of the Weyl tensor can then act as the microstates of the gravitational field, and the entropy would from coarse-graining over states of the Weyl tensor. The counting of allowed microstates would be influenced by the boundary conditions imposed by the local stress-energy. The cosmological horizon suppresses some long-wavelength modes of the Weyl tensor, leading the entropy to be lower for a larger cosmological constant. Likewise, the presence of a black hole or a star changes which modes are allowed and can depress the entropy.

\section{The End State of the Universe}
The equilibrium state of a gas in a box is a state of maximum entropy where the gas is uniformly distributed and featureless. The universe today clearly is not in gravitational equilibrium. The process of objects passing beyond the cosmological horizon, and of black holes evaporating, are the processes by which the universe is reaching equilibrium. Empty de Sitter space is maximally symmetric since it posses a full set of ten symmetries expressed by Killing vectors (one time and three space translations, three rotations, and three boosts). It is the space with maximum entropy of the spacetimes with fixed $\Lambda$ \cite{ref27,ref28}. Flat space has the same symmetries, but does not have the preferred scale that $\Lambda$ provides. It therefore has even higher entropy than de Sitter.  When quantum effects are included de Sitter may decay to flat space. Alternatively, if the cosmological constant is mimicked by a scalar field, our universe could cascade through a series of configurations with increasing entropy and decreasing apparent dark energy density, and eventually reach flat space. All of these processes are consistent with the second law of thermodynamics, as entropy continually increases.

\section{The Entropy of the Gravitational Field}
A completely satisfactory theory of gravitational thermodynamics is very challenging because energy at a point cannot be defined in general relativity due to the equivalence principle - going into free fall removes all local gravitational effects. One solution to this problem is quasilocal energy where one computes the energy inside a surface. The universal nature of gravitation means that the quasilocal energy counts all of the energy inside a surface, not just gravitational energy. As an example, we will derive formulas for entropy associated with a star using Brown-York quasilocal quantities \cite{BrownYork,LundgrenBrownYork}. We start with the usual expression of the first law of thermodynamics 
\begin{equation}
\mathrm{d}U = T~\mathrm{d}S - p~\mathrm{d}V ~,
\end{equation}
relating the change in internal energy $U$ to changes in the entropy $S$ and volume $V$ through the temperature $T$ and pressure $p$. This law does not extend naturally to gravitating system because $U$, $S$, and $V$ are extensive variables. When two subsystems are brought together, the extensive quantities should add. However, this does not work for unshielded long-range forces like gravitation.  As two stars are brought together the gravitational binding energy deceases the total energy.  But quasilocal energy is additive since when  two regions share a common boundary the contributions to the two quasilocal integrals cancel.

Similarly, gravitation curves space so the total volume depends in a nonlinear way on the overall mass distribution. For a surface enclosing a black hole, the volume inside is not really a sensible quantity. Quasilocal quantities suggest a replacement for the $p \mathrm{d}V$ term with $s \mathrm{d}A$, where $s$ is the surface pressure of \cite{BrownYork}, which is perhaps better thought of as a surface tension. We can stretch the quasilocal surface, but we suffer a change in energy proportional to the surface tension times the change in area. This is a much more natural quantity in curved space than a change in volume.

To demonstrate the use of these quantities, we consider a static, spherically-symmetric star with density $\rho(r)$ and pressure $p(r)$. The metric is
\begin{equation}
ds^2 = - N^2(r) dt^2 + h^2(r) dr^2 + r^2 d\Omega^2 ~.
\end{equation}
Solving the Einstein equation determines both $h(r)$ and $N'(r)/N(r)$ in terms of $\rho(r)$ and $p(r)$. In what follows, we will also use the ``mass'' of the star,
\begin{equation}
m(r) \equiv 4 \pi \int_0^r \bar{r}^2 \rho(r) \mathrm{d}r + M ~.
\end{equation}
The constant $M$ allows for the possibility of a black hole rather than a star.

The quasilocal energy is
\begin{equation}
E = - \frac{r}{h(r)} = - r \sqrt{1 - 2 m(r) / r}
\end{equation}
and the quasilocal surface tension is
\begin{equation}
s = \frac{1}{8 \pi} \Big[ \frac{N'(r)}{N(r) h(r)} + \frac{1}{r h(r)} \Big] = \frac{h(r)}{8 \pi r^2} \Big[ r - m(r) + 4 \pi r^3 p(r) \Big] ~.
\end{equation}
The first law of thermodynamics becomes 
\begin{equation}
\mathrm{d}E = T \mathrm{d}S - s \mathrm{d}A.
\end{equation}
 Note that both $E$ and $s$ include a subtraction term, which sets the value of each to $0$ in flat space (this is obtained by setting $h(r) = N(r) = 1$), but these cancel. We allow two kinds of virtual displacements - changes in the radius of the quasilocal surface and in the mass profile of the star.
\begin{align}
T~\mathrm{d}S &= \mathrm{d}E + s~\mathrm{d}A \\
& = \Big( \rho(r) + p(r) \Big)~\mathrm{d}V + h(r) ~ \mathrm{d}m(r)
\end{align}
where $\mathrm{d}V = 4 \pi h(r) r^2 \mathrm{d}r$ is the curved space volume.

The first term is zero when $p = - \rho$, so the entropy is constant outside the star even when there is a cosmological constant. However, inside the star, the entropy is not constant. The temperature of the cosmological horizon is $T_C \sim (2 \pi \alpha)^{-1}$. So if the total mass of the star is $M_s$, then the entropy at the center of the star is less than that at the surface by an amount proportional to $M_s \alpha$.

We have pointed out that a simple problem like a star or black hole in de Sitter space is far from equilibrium. Any gravitational configuration other than de Sitter space may be best treated as a non-equilibrium state. We suggest to replace the $E$ and $p~\mathrm{d}V$ terms with quasilocal quantities. A complete treatment of gravitational entropy involves dealing with the long-range and non-equilibrium nature of the gravitational field. The positive cosmological constant plays a large role in the thermodynamics of the gravitational field, and better understanding of its role will improve our understanding of the evolution of the universe and perhaps of cosmology as a whole. 

{\it Acknowledgements.}  
APL is grateful to James York Jr. for introducing him to black hole thermodynamics. RB is supported by the Dr. Tomalla Foundation and the Swiss National Foundation. We thank Alex Nielsen, Badri Krishnan, Bjoern Schmekel, and Philippe Jetzer for useful discussions, encouragement and support. In particular, we thank Alex Nielsen for pointing out that the black hole can be replaced by a spherical star with the same mass.  We thank our session chair from MG13 for emphasizing that the decay to flat space of deSitter would be consistent with the second law of thermodynamics. 

\end{document}